\documentclass[%
 reprint,
 amsmath,amssymb,
 aps,
]{revtex4-2}

\usepackage{graphicx}
\usepackage{dcolumn}
\usepackage{bm}
\usepackage{hyperref}
\usepackage{xcolor}
\usepackage{subcaption}  

\usepackage{soul}
\usepackage{xcolor}

\begin{document}

\preprint{APS/123-QED}

\title{A generative machine learning surrogate model of plasma turbulence.}
\author{B. Clavier}
 \email{benoit.clavier@univ-amu.fr}
\author{D. Zarzoso}
\affiliation{Aix Marseille Univ, CNRS, Centrale Med, M2P2 UMR 7340, Marseille}
\author{D. del-Castilllo-Negrete}
\affiliation{Oak Ridge National Laboratory, Oak Ridge, TN 37831-8071, United States of America }
\author{E. Fr\'enod}
\affiliation{Universit\'e Bretagne Sud, LMBA UMR 6205, Vannes}

\begin{abstract}
Generative artificial intelligence methods are employed for the first time to construct a surrogate model for plasma turbulence that  enables long time transport simulations.
The proposed GAIT (Generative Artificial Intelligence Turbulence) model is based on the coupling of a convolutional variational auto-encoder, that encodes precomputed turbulence data into a reduced latent space, and a recurrent neural network and decoder that generates new turbulence states 400 times faster than the direct numerical integration. The model is applied to the Hasegawa-Wakatani (HW) plasma turbulence model, that is closely related to the quasigeostrophic model used in geophysical fluid dynamics. Very good agreement is found between the GAIT and the HW models in the spatio-temporal Fourier and Proper Orthogonal Decomposition spectra, and the flow topology characterized by the Okubo-Weiss decomposition. The GAIT model also reproduces Lagrangian  transport including the probability distribution function of particle displacements and the effective turbulent diffusivity.
\end{abstract}

\maketitle

Turbulence is ubiquitous in nature and industrial systems. The atmosphere of planets \cite{wyngaard1992atmospheric}, oceanic currents \cite{gargett1989ocean}, ionized gases in stars \cite{canuto1998turbulence}, the solar tachocline \cite{spiegel1992solar}, and the solar wind \cite{bruno2013solar} are some examples of systems where turbulence can be encountered.
Without doubt, turbulence represents one of the greatest challenges in numerical modeling due to the vast range of spatiotemporal scales involved and the unpredictable behavior.

In this Letter we focus on turbulence in magnetized plasmas of interest to controlled nuclear fusion and astrophysics. 
The pressing need to understand and predict the role of turbulence in particles and energy transport in fusion plasmas has motivated the development of a large number of computational tools ranging from fluid to gyro-fluid and gyro-kinetic models. However, due to the spatiotemporal multiscale properties of plasma turbulence these computations are very time consuming and in some cases impractical or unaffordable. To overcome this limitation we propose the use of state-of-the-art artificial intelligence (AI) methods to accelerate turbulence simulations. 
Although our focus is on plasma physics, the turbulence model of interest contains as a special case a model extensively used in geophysical fluid dynamics.
Recently, machine learning techniques have been used to accelerate and validate plasma physics simulations as well as to develop data-driven surrogate models \cite{pathak2018model,zhu2022data,rodriguez2018vitals,churchill2023accelerating, greif2023physics,Yang:2023,mcdevitt2023physics,pathak2018model}.
Of particular interest is the potential of AI to create new data from existing data using generative deep-learning techniques. These techniques, which have been used in chemistry to automatically design new molecules \cite{gomez2018automatic} and fluid mechanics \cite{wang2024towards}, form the basis of our proposed GAIT (Generative Artificial Intelligence Turbulence) surrogate model. 

Once computed, a turbulence field can be used to perform transport studies of passive tracers, e.g., impurities. In the simplest setting, the basic idea is to solve the transport equation in a time interval $(0,t_{max})$ for a large ensemble of particles using the electromagnetic fields computed from the numerical solution of a turbulence model (e.g., gyro-kinetics or gyro-fluid). The main limitation encountered in this type of studies is that the $t_{max}$ of relevance for transport studies is typically much larger than the time range for which the turbulence model can be solved, due to limited computational resources.  One way to overcome this limitation is by using surrogate models, like the GAIT model proposed here, that once trained in a time range $(0,t_{train})$ can be used to produce, with negligible computational cost, turbulent states in the transport scale range $(0,t_{max})$ where $t_{train} \ll t_{max}$.

The fast production of  long-time turbulent simulations are also useful for capturing the correct asymptotic behavior of tracers exhibiting rare events in the Lagrangian statistics.  Examples include anomalously large trapping and/or flights \cite{del1998asymmetric}, and anomalous diffusion transitions beyond intermediate asymptotic regimes \cite{cartea2007fluid}.

As a tractable example to illustrate and test the GAIT model we consider the extensively used 
Hasegawa-Wakatani (HW) model \cite{hasegawa1983plasma} 
\begin{eqnarray}
\partial_t n + \left[\phi,n\right] & = & C\left(\phi - n\right) -\kappa\partial_y\phi + D_c \nabla^2n \label{HW_n}\\
\partial_t \Omega + \left[\phi,\Omega\right] & = & C\left(\phi - n\right) + \nu\nabla^2\Omega \label{HW_Omega}
\end{eqnarray}
where $\phi$ is the electrostatic potential, $n$ is the density, and $\Omega=\nabla^2_\bot\phi$ the vorticity in the $(x,y)$ plane. $C$ is the adiabaticity coefficient, $D_c$ is the density diffusivity, $\nu$ is the viscosity, $\left[A_1,A_2\right]=\partial_xA_1\partial_yA_2 - \partial_yA_1\partial_xA_2$ is the Poisson bracket, and $\kappa=-\partial_x\log n_0$ is the drive of the instability underlying the turbulence with $n_0$ the background density. The magnetic field is assumed of the form $\mathbf{B}=B\mathbf{e}_z$ with $B$ constant.
Time is normalized to the cyclotron frequency $\omega_{c0}$, and space to the Larmor radius $\rho_s$. In applications to controlled nuclear fusion in toroidal magnetic confinement devices, this model is used to describe edge turbulence and recently used in   \cite{gahr2024learning} for data-driven turbulence modeling.

Assuming $D_c=\nu=0$, in the limit $C\rightarrow \infty$, $\phi=n$ and Eqs.(1)-(2) reduce to the Charney-Hasegawa-Mima equation describing Rossby waves in geophysical flows \cite{pedlosky2013geophysical} and drift waves in inhomogeneous plasmas \cite{hasegawa1977stationary}. In this reduction, the magnetic field corresponds to the Earth's rotation, the  potential to the fluid streamfunction, and the plasma density gradient to the variation of the Coriolis force with latitude. This analogy allows the application of methods and ideas in magnetized plasmas to geophysical flows and viceversa, see for example Ref.\cite{del2000chaotic} and references therein. Since our methodology focuses on the construction of a surrogate model for the potential, and not the density, it is directly applicable to  the study of turbulence and transport in geophysical flows \cite{bracco2004dispersion}.

\begin{figure}[h]
    \includegraphics[width=\linewidth]{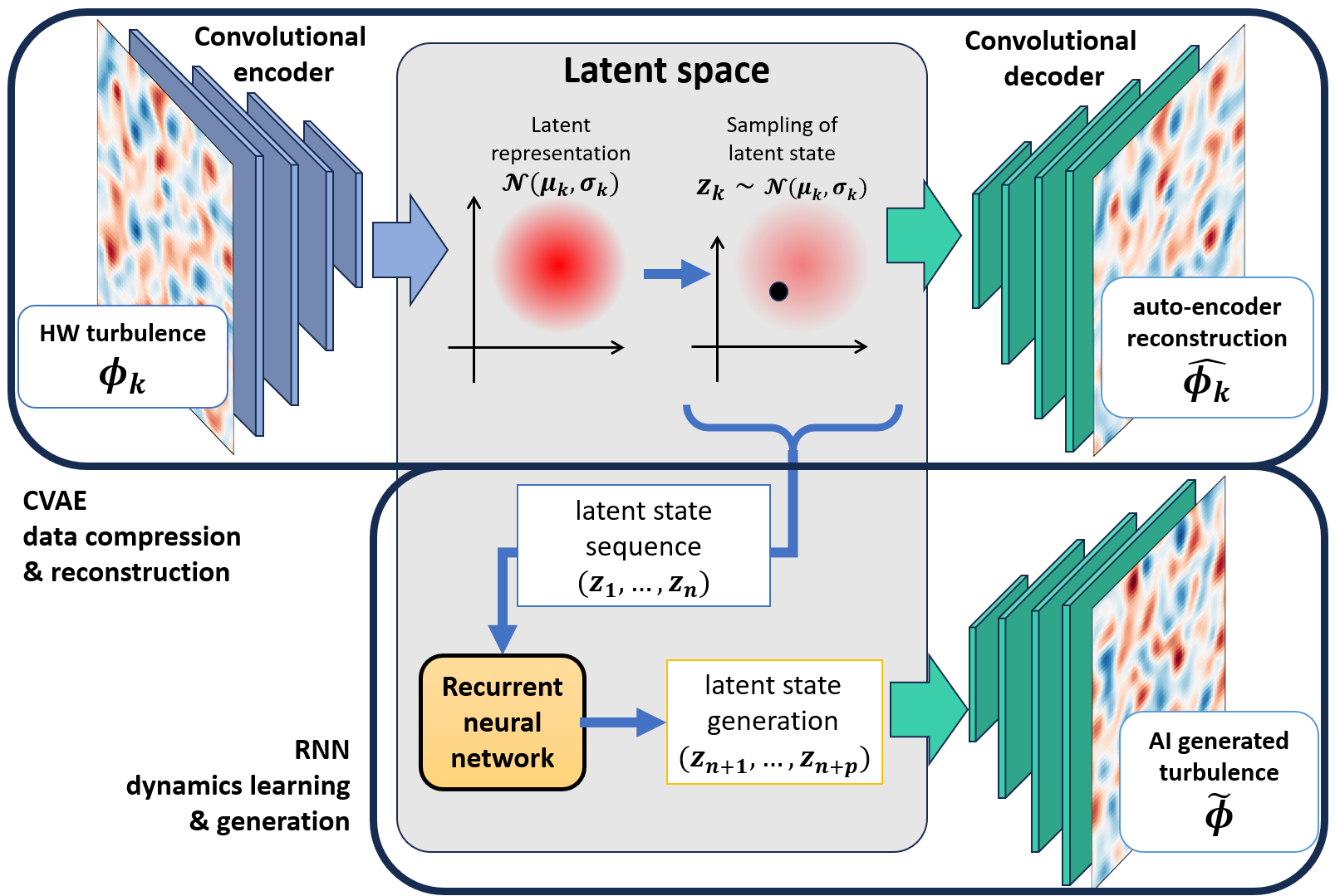}
    \caption{\raggedright \label{fig:CVAE-RNN} Schematic representation of the architecture of the proposed GAIT (Generative Artificial Intelligence Turbulence) model. [For further details see supplemental material \cite{supplemental}]
}
\end{figure}

To train the GAIT model, Eqs.(\ref{HW_n})-(\ref{HW_Omega}) with $C=1$, $\kappa=1$, $D_c=\nu=10^{-3}$ and $k_0=0.15$ were solved numerically using a pseudo-spectral method with a $\left(N_x,N_y\right)=512\times 512$ spatial grid and a 4th order Runge-Kutta scheme for the time advance with $\Delta t = 2\cdot 10^{-2}\omega_{c0}^{-1}$ to guarantee convergence and stability. The physical parameters are within the range of what is typically used in the literature \cite{anderson2020elucidating} \cite{pushkarev2013self}, and the resolution guarantees the numerical stability of the simulation \cite{greif2023physics}. 

For the size of the training data we considered $N_s=8000$ snapshots of the potential obtained by saving the  data at a rate $\Delta t_{saving} = \omega_{c0}^{-1}$, i.e. every 50 numerical steps. To eliminate transients, snapshots were saved after the turbulence saturation time $\omega_{c0} t \simeq 200$. Information on the dependence of the  GAIT model on $N_s$ can be found in the supplemental material \cite{supplemental}.

\begin{figure}[h]
    \begin{subfigure}{0.135\textwidth}
        \centering
        \includegraphics[height=65pt, trim=40 10 70 35, clip]{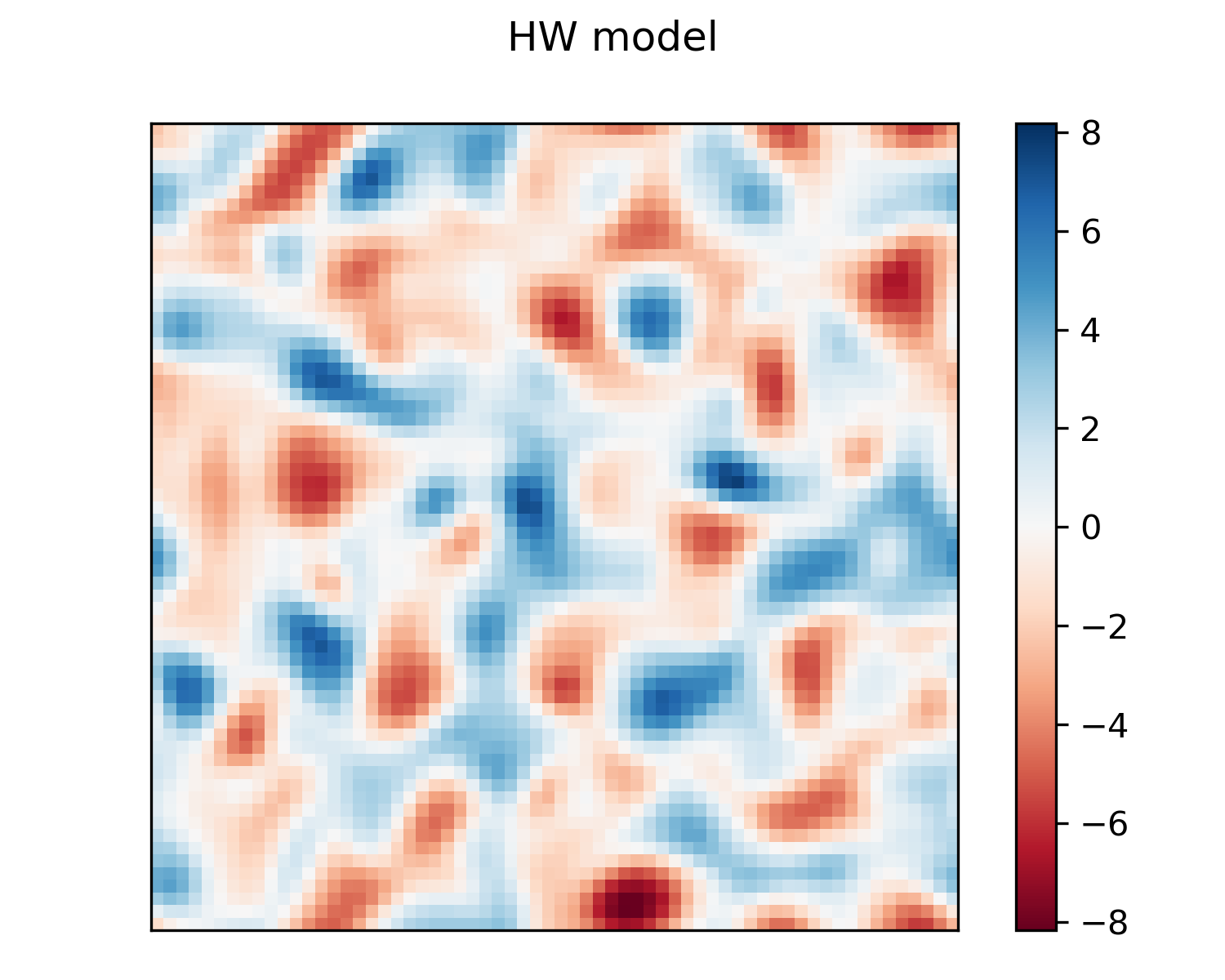}
        \label{fig:subfig01}
    \end{subfigure}
    \begin{subfigure}{0.135\textwidth}
        \centering
        \includegraphics[height=65pt, trim=40 10 20 35, clip]{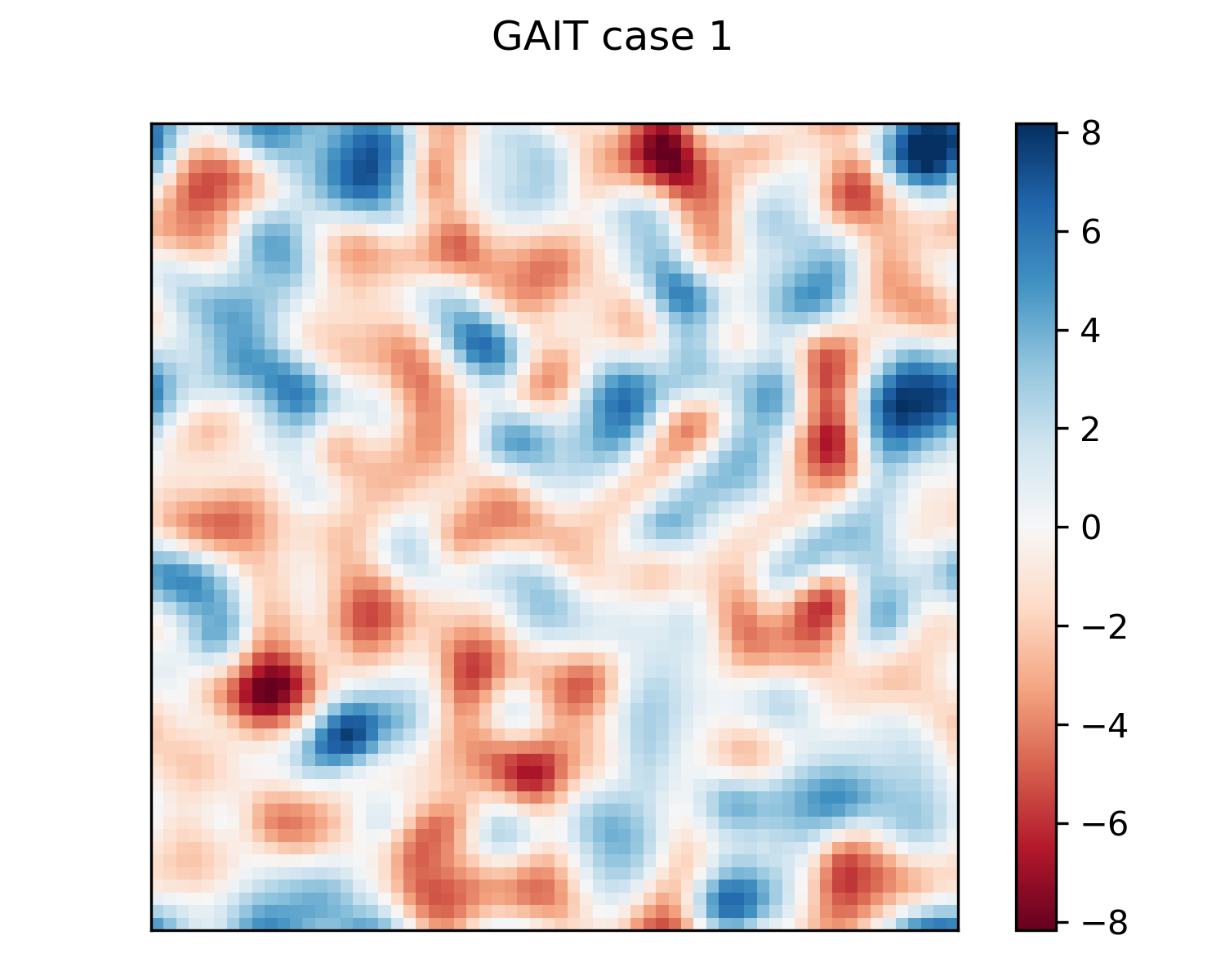}
        \label{fig:subfig01}
    \end{subfigure}
    \caption{\raggedright Time snapshots of electrostatic potential, $\phi$, in $(x,y)$ obtained from the direct numerical integration of the HW (Hasegawa-Wakatani) model (left) and the proposed GAIT model.}
    \label{fig:eulerian2}
\end{figure}

As illustrated in Fig.~\ref{fig:CVAE-RNN}, the proposed GAIT model consists of a convolutional variational auto-encoder (CVAE) coupled to a
recurrent neural network (RNN), a detailed animation can be found in the supplemental material \cite{supplemental}. 
CVAEs are probabilistic generative machine learning models composed of an encoder, ${\cal E}$,  and a decoder, ${\cal D}$ \cite{kingma2013auto, cheng2020advanced}. A similar architecture has been also used to compress diagnostic measurements in plasmas \cite{zhu2022data}. 
In our problem  ${\cal E}$ is a convolutional neural network (CNN) 
consisting of five successive 2D convolution layers (with variable stride, kernel sizes, and filters)
that map snapshots of the electrostatic potential, $\phi$, represented in an $N_x \times N_y$ grid of the $(x,y)$ plane, to points 
in a reduced $N$-dimensional latent space of parameters of a variational Gaussian distribution $\mathcal{N}\left(\mathbf{z}\vert\mathbf{\mu},\mathbf{\sigma}\right)$, with $\mathbf{z}\in\mathbb{R}^N$, $\mathbf{\mu}\in\mathbb{R}^N$ and $\mathbf{\sigma}\in\mathbb{R}^N$. 
For training purposes, the $512 \times 512$ turbulence data was down-sampled to a $N_x \times N_y = 64 \times 64$ resolution. For the dimension of the latent space we used $N=64$, resulting in a encoder data compression factor of $N_x N_y/N=64$.
The decoder, ${\cal D}$, 
is designed as a mirror image of the encoder with transposed convolution layers that progressively unfold $N$-dimensional latent vectors into $N_x \times N_y$ dimensional images, such that 
${\cal D} [{\cal E}[\phi]]=\hat \phi \approx \phi$.
The goal of the training is to find an optimal ${\cal E}$ that compresses the data preserving the maximum information 
and an optimal ${\cal D}$ that has the minimum reconstruction error. This is achieved by minimizing the loss function
\begin{equation}\label{loss_function_CVAE}
\begin{split}
L_{\rm CVAE} = & \sum_i^{N_s} w_\phi\left\Vert\phi_i-\hat{\phi_i}\right\Vert^2 + w_{\nabla\phi}\left\Vert\nabla\phi_i-\nabla\hat{\phi_i}\right\Vert^2 \\
& + w_{\mathbb{K}\mathbb{L}}\mathbb{K}\mathbb{L}\left(\mathcal{N}\left(\mathbf{z}_i\vert\mathbf{\mu}_i,\mathbf{\sigma}_i\right),\mathcal{N}\left(\mathbf{z}_i\vert\mathbf{0},\mathbf{1}\right)\right)
\end{split}
\end{equation}
where $\{\phi_i\}_{=1}^{N_{s}}$ and $\{\hat \phi_i\}_{=1}^{N_{s}}$ are the $N_s$ training and the $N_s$ reconstructed snapshots respectively, $\mathbf{z}_i=\mathcal{E}\left(\mathbf{\phi}_i\right)$.
The first two terms in Eq.~(\ref{loss_function_CVAE}) are the reconstruction loss of $\phi$ and $\nabla \phi$ in the L2 norm, weighted by the hyper-parameters  $w_\phi\geq 0$ and  $w_{\nabla \phi}\geq 0$. 
The third term in Eq.(\ref{loss_function_CVAE}), with weight $w_{\mathbb{K}\mathbb{L}}\geq 0$, uses the Kullback-Leibler divergence $\mathbb{K}\mathbb{L}$ \cite{kullback1951information} to quantify the difference between the normal distributions of the mapped snapshots in the latent space with mean $\mu_i$ and variance $\sigma_i$ and a normal distribution with zero mean and unit variance. Minimizing this difference forces the encoder to map the different images into a compact set with small variability in the latent space, a property that is critical for the success of the generation of turbulence for transport studies. 

Since $L_{\rm CVAE}$ is defined up to a scale factor, there are only two independent parameter ratios, which for $w_{\mathbb{K}\mathbb{L}}\neq 0$ must satisfy $w_\phi/w_{\mathbb{K}\mathbb{L}}>1$ (if $w_\phi\neq 0$) and  $w_{\nabla \phi}/w_{\mathbb{K}\mathbb{L}}>1$ (if $w_{\nabla \phi}\neq 0$), so that the $\mathbb{K}\mathbb{L}$ loss does not dominate over the reconstruction loss. Increasing the value of a hyper-parameter will yield a higher accuracy in the observable(s) associated to it. For example, increasing $w_{\nabla \phi}$ will increase the accuracy of observables like the electric field ${\bf E} =- \nabla \phi$ and the ${\bf E} \times {\bf B}$ velocity.  $w_{KL}$ controls the regularity of the latent space. Decreasing this parameter too much leads to noisy Lagrangian dynamics that might artificially increase the diffusivity.

The training of the CVAE was implemented using an Adam optimizer with a learning rate $l_r=10^{-3}$ and 5000 epochs. Since the numerical simulations are performed in a double periodic domain, the  convolution algorithms of TensorFlow were modified so the encoder captures the periodicity of the input data and the decoder generates periodic images from points in the latent space.

Following the  CVAE  training, the sequence  $\{{\bf z}_1,{\bf z}_2, \ldots {\bf z}_{N_s}  \}$, corresponding to the evolution in time of the turbulence snapshots $\{\phi_1(x,y),\phi_2(x,y), \ldots {\phi}_{N_s}(x,y)  \}$, is split into training and testing sequences containing 85\% and 15\% of the data respectively.
The training sequence is then used to train a multi-layer RNN (Recurrent Neural Network) with two hidden recurrent layers,  using an Adam optimizer with learning rate $l_r=10^{-3}$, 500 epochs, and loss function  $L_{\rm RNN} = \sum_i \left\Vert({\bf z}_{i+1}, \ldots {\bf z}_{i+l+1})-RNN({\bf z}_{i}, \ldots {\bf z}_{i+l})\right\Vert^2$ with $l = 50$ to correctly reproduce the time auto-correlation.
Once trained, the RNN is used to advance the  latent space sequence from $\{{\bf z}_{i}, \ldots {\bf z}_{i+l}  \}$ to $\{{\bf z}_{i+1}, \ldots {\bf z}_{i+l+1}  \}$.

The turbulence AI generation process is completed by using the decoder to map the new points in the latent space into new snapshots of turbulence, $\{\tilde \phi_{N_s+1}(x,y),\ldots \tilde {\phi}_{N_s+N_g}(x,y)  \}$ with $\tilde \phi_{j}(x,y)={\cal D}[{\bf \tilde z}_j]$.
As shown in Fig.\ref{fig:eulerian2}, the images of the AI generated turbulent states are practically indistinguishable from those generated from the direct numerical simulation of the Hasegawa-Wakatani turbulence model in Eq.~(\ref{HW_n})-(\ref{HW_Omega}). However, from the physics perspective it is critical to go beyond the observed qualitative agreement of the images and perform systematic tests using quantitative metrics. The rest of the letter is devoted to this goal.

We present numerical results for two realizations of the GAIT model. \textit{Case 1}, with $(w_\phi, w_{\nabla \phi}, w_{\mathbb{K}\mathbb{L}}) = (1, 10, 0.01)$, optimizes the reconstruction of $\phi$ and $\nabla \phi$, with low latent space regularity. \textit{Case 2}, with $(w_\phi, w_{\nabla \phi}, w_{\mathbb{K}\mathbb{L}}) = (0, 10, 0.1)$, optimizes only $\nabla \phi$, with high latent space regularity.

\begin{figure}[h]
    \begin{subfigure}{0.23\textwidth}
        \centering
        \includegraphics[width=0.9\linewidth, trim=0 0 0 0, clip]{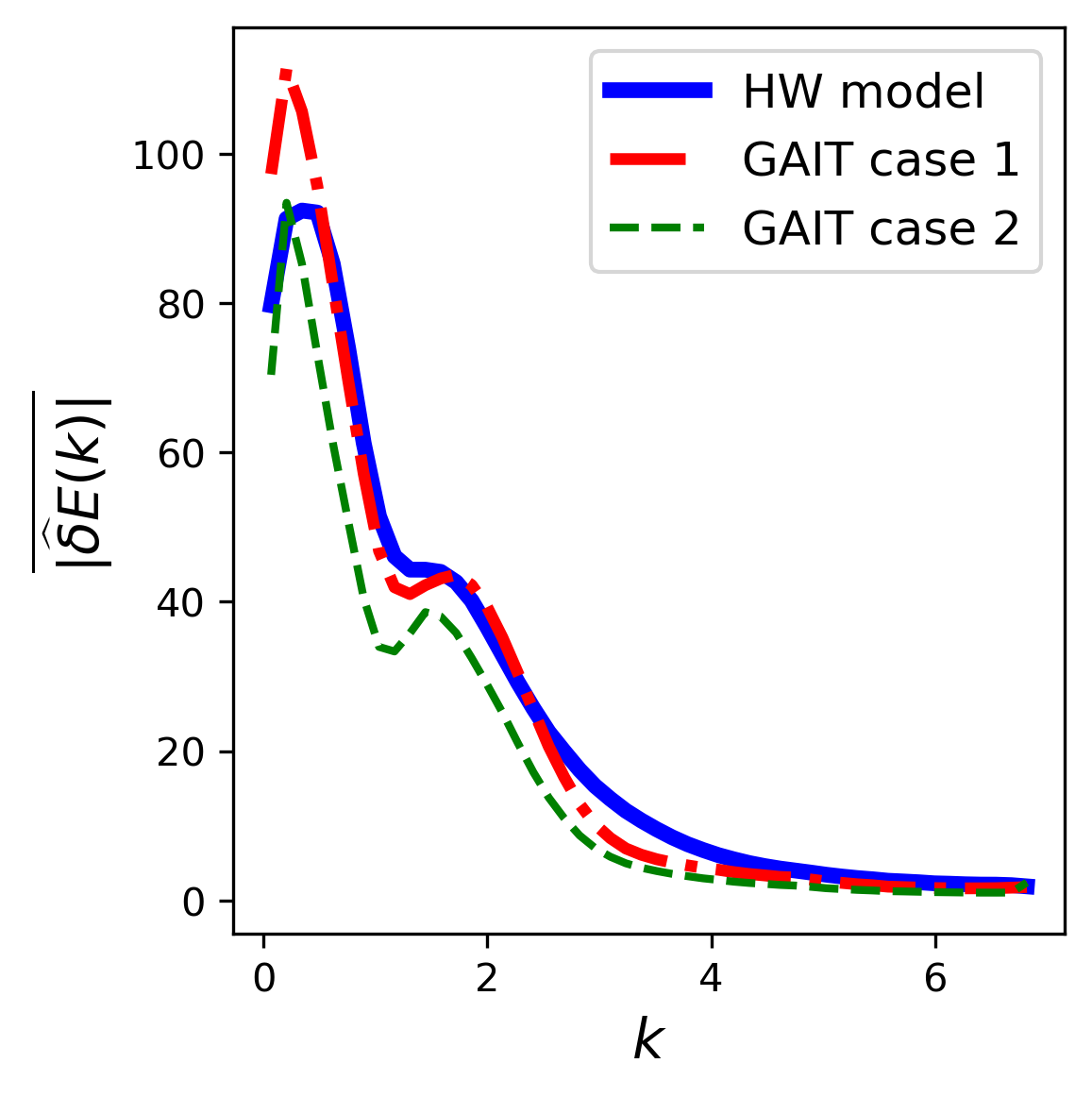}
        \label{fig:subfig1}
    \end{subfigure}
    \begin{subfigure}{0.23\textwidth}
        \centering
        \includegraphics[width=0.9\linewidth, trim=0 0 0 0, clip]{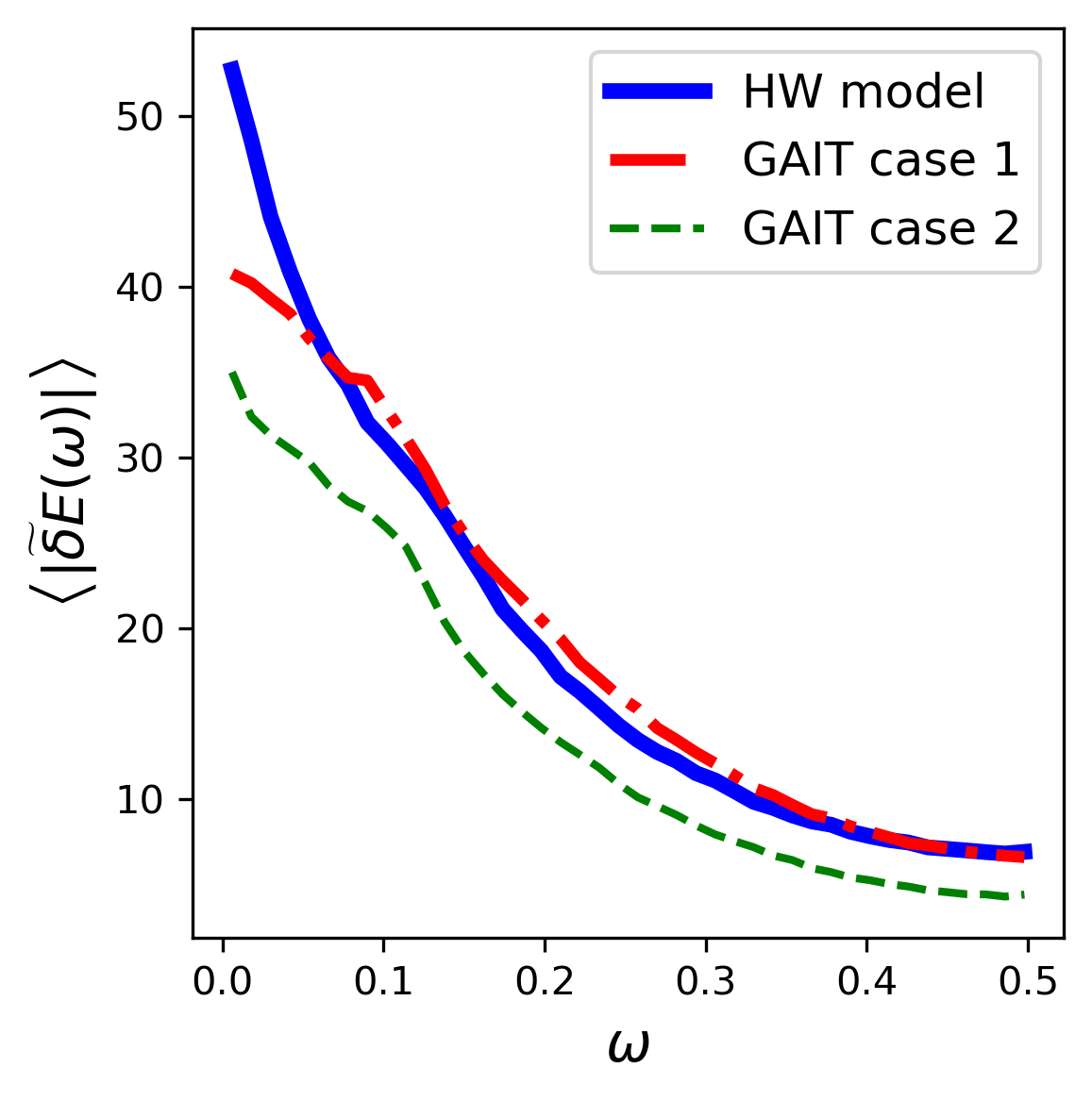}
        \label{fig:subfig2}
    \end{subfigure}
    \caption{\raggedright Comparison between the HW and GAIT models electric field turbulent fluctuations power spectra in space (left) and time (right).
}
    \label{fig:eulerian1}
\end{figure}

The first test compares the Fourier power spectra in space of the electric field turbulent fluctuations,
$\overline{|\widehat \delta E (k)|}$, where $\widehat \delta E (k,t)$ is the Fourier transform in space of $\delta E(x,y,t)= E(x,y,t) - \left< E\right>$, 
$\left< f \right>$ denotes spatial average, $\overline f $ denotes time average,
$E$ is the magnitude of the electric field, $E=| -\nabla \phi |$, and $k=\sqrt{k_x^2+k_y^2}$.
The second test compares the Fourier power spectra in time of the electric field turbulent fluctuations,
$\left < |\widetilde \delta E (\omega)| \right>$,
where $\widetilde \delta E (x,\omega)$ is the Fourier transform in time of $\delta E(x,y,t)$. Figure~\ref{fig:eulerian1} compares these two metrics in the HW and the GAIT models.

\begin{figure}[h]
    \begin{subfigure}{0.23\textwidth}
        \centering
        \includegraphics[width=0.9\linewidth, trim=0 0 0 0, clip]{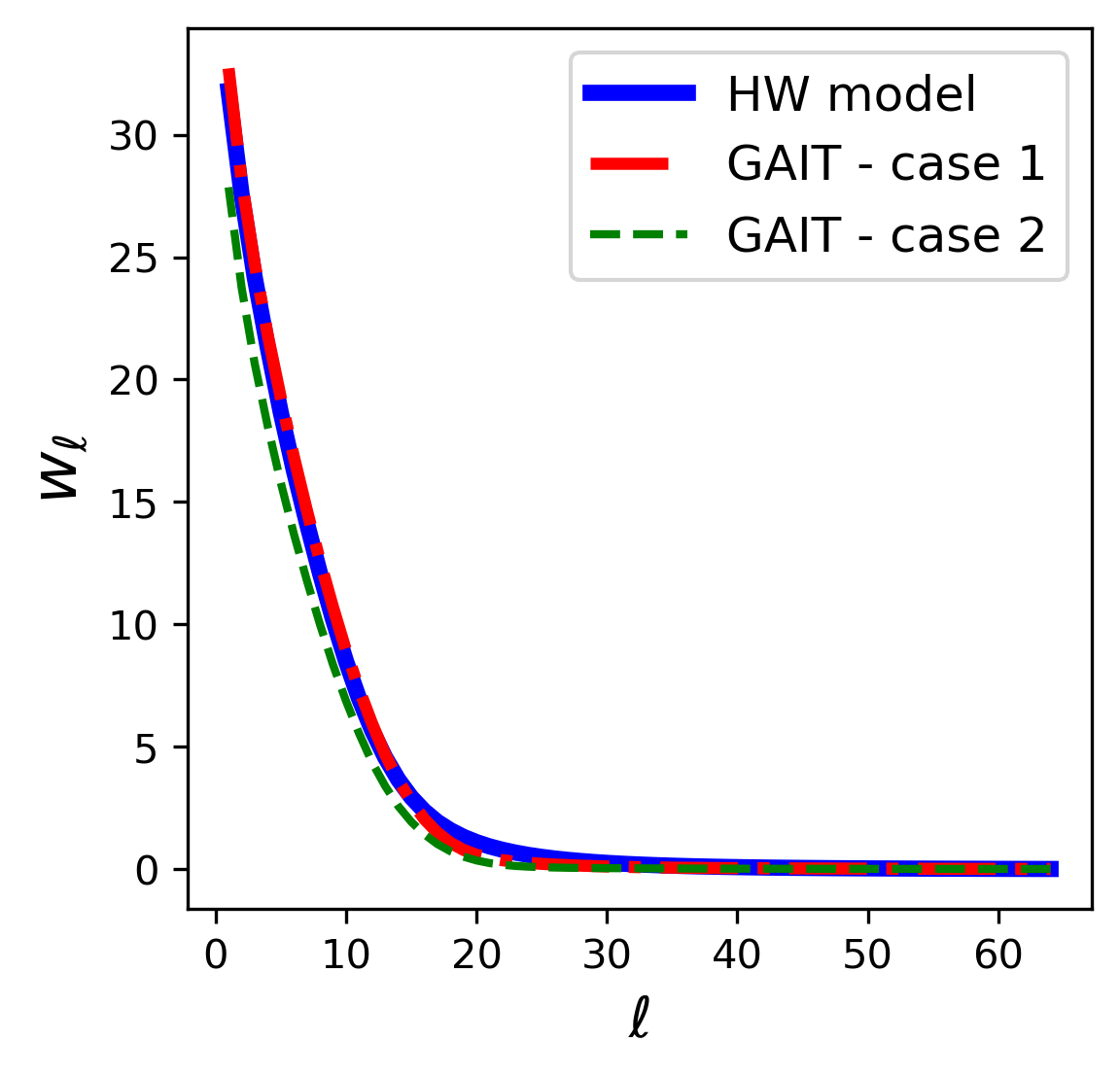}
        \label{fig:subfig3}
    \end{subfigure}
    \begin{subfigure}{0.23\textwidth}
        \centering
        \includegraphics[width=0.9\linewidth, trim=0 0 0 0, clip]{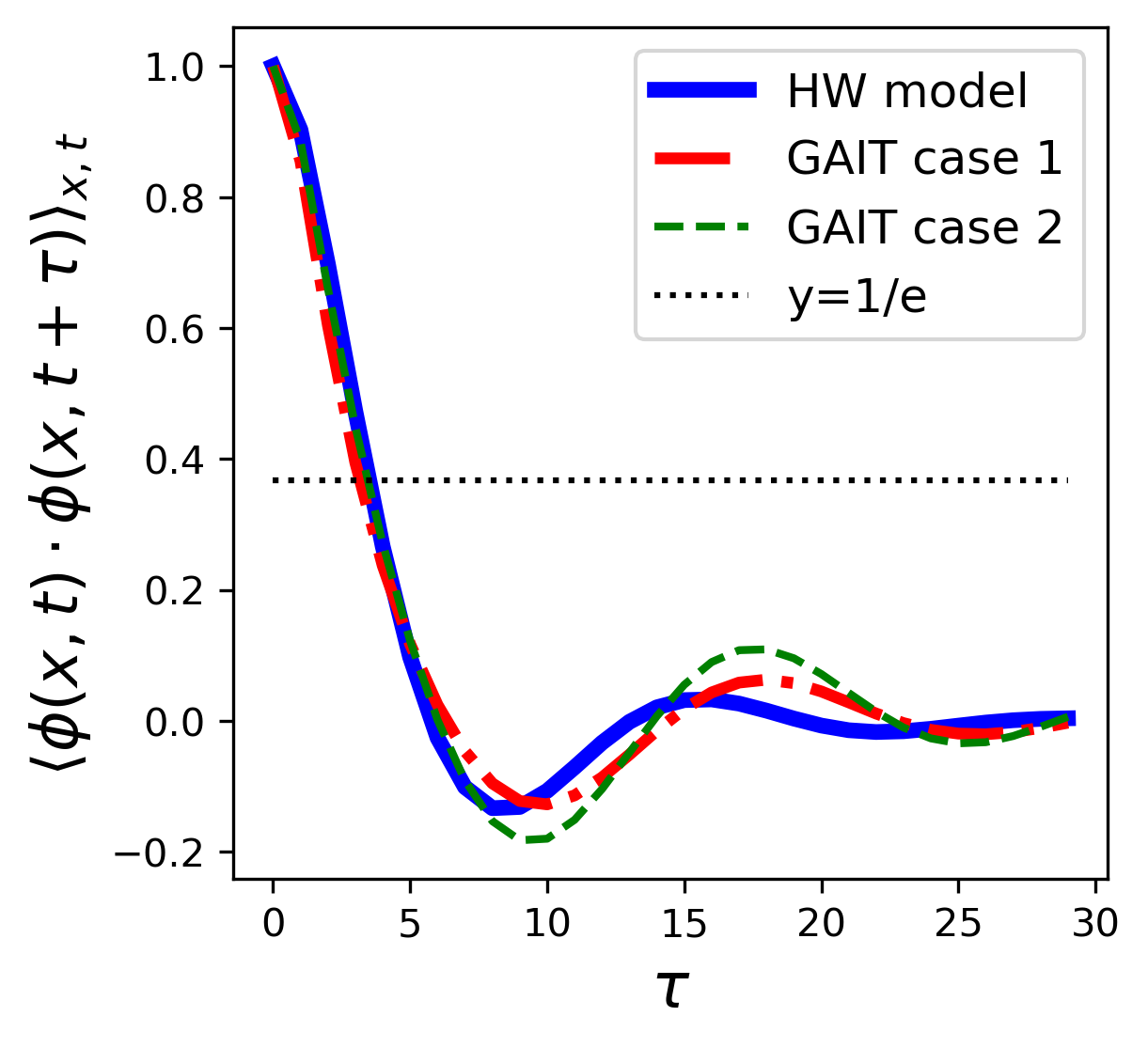}
        \label{fig:subfig4}
    \end{subfigure}
    \caption{\raggedright 
    Comparison between the HW and GAIT models proper orthogonal decomposition spectrum of potential, $\phi$, (left) and time auto-correlation function (right).
    }
    \label{fig:eulerian3}
\end{figure}

\begin{figure}[h]
    \begin{subfigure}{0.15\textwidth}
        \centering
        \includegraphics[height=65pt, trim=0 0 0 20, clip]{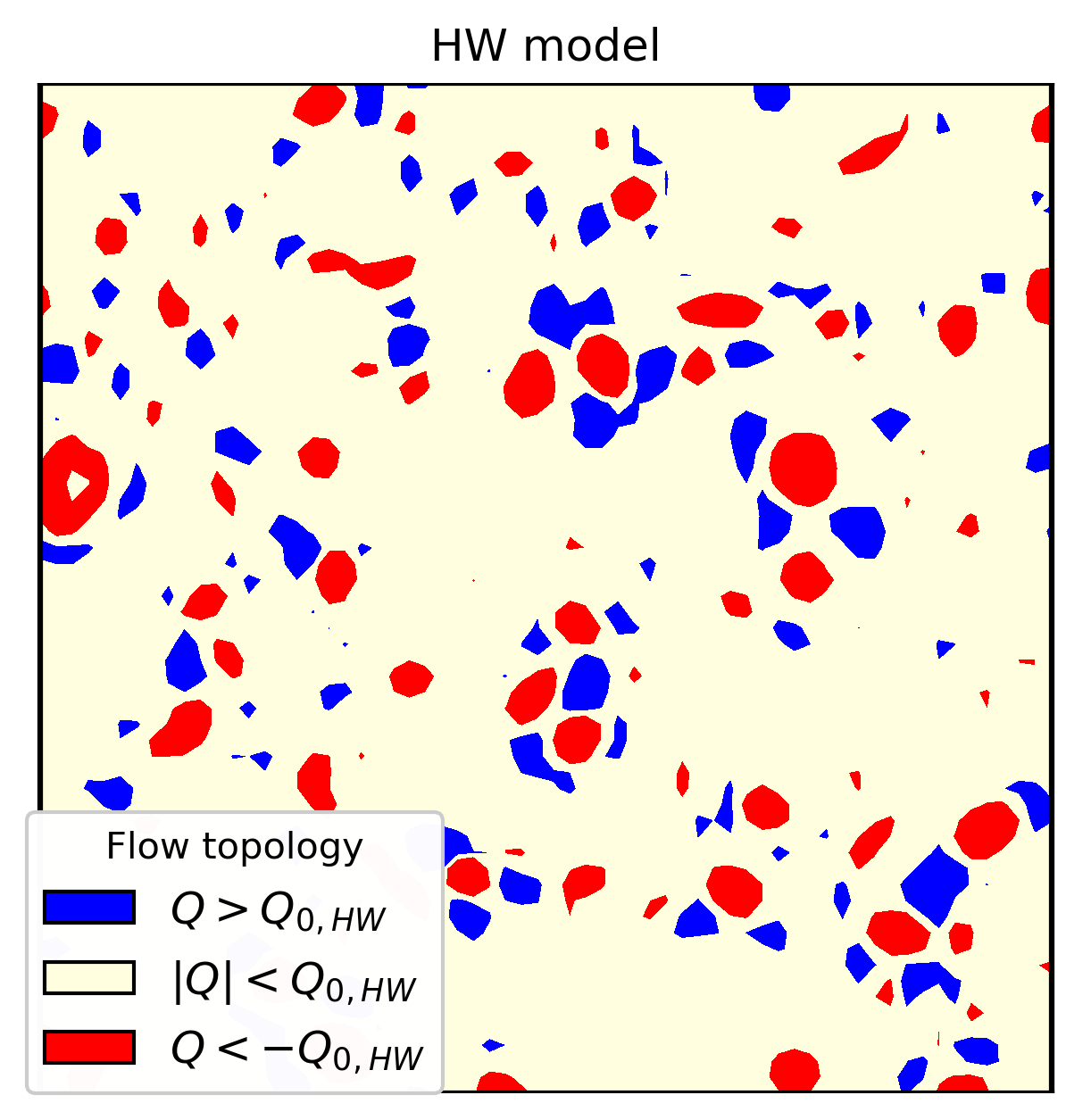}
        \label{fig:subfig01}
    \end{subfigure}
    \begin{subfigure}{0.15\textwidth}
        \centering
        \includegraphics[height=65pt, trim=0 0 0 20, clip]{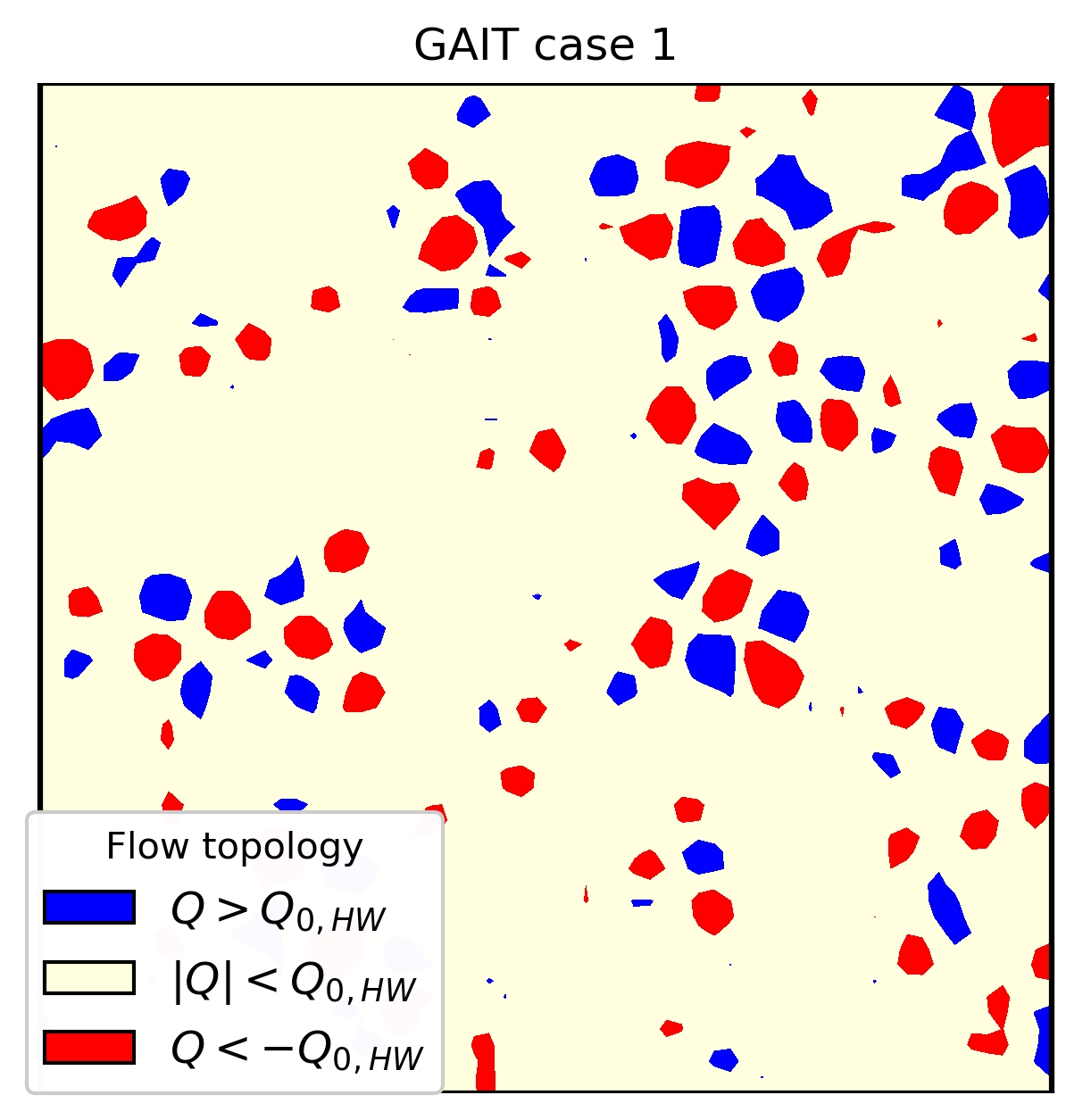}
        \label{fig:subfig01}
    \end{subfigure}
    \begin{subfigure}{0.15\textwidth}
        \centering
        \includegraphics[height=65pt, trim=0 0 0 20, clip]{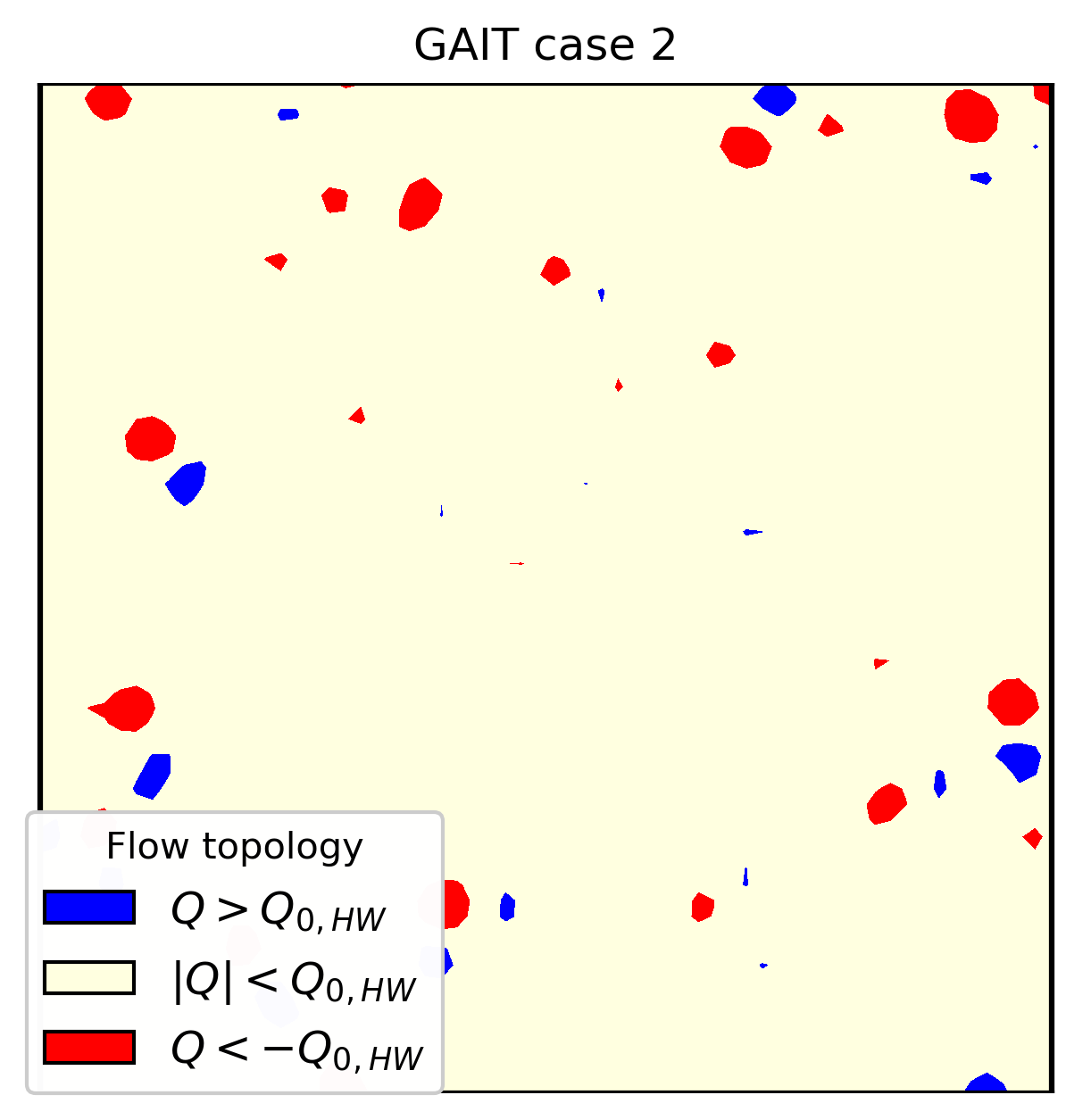}
        \label{fig:subfig01}
    \end{subfigure}
    
     \includegraphics[scale=0.3]{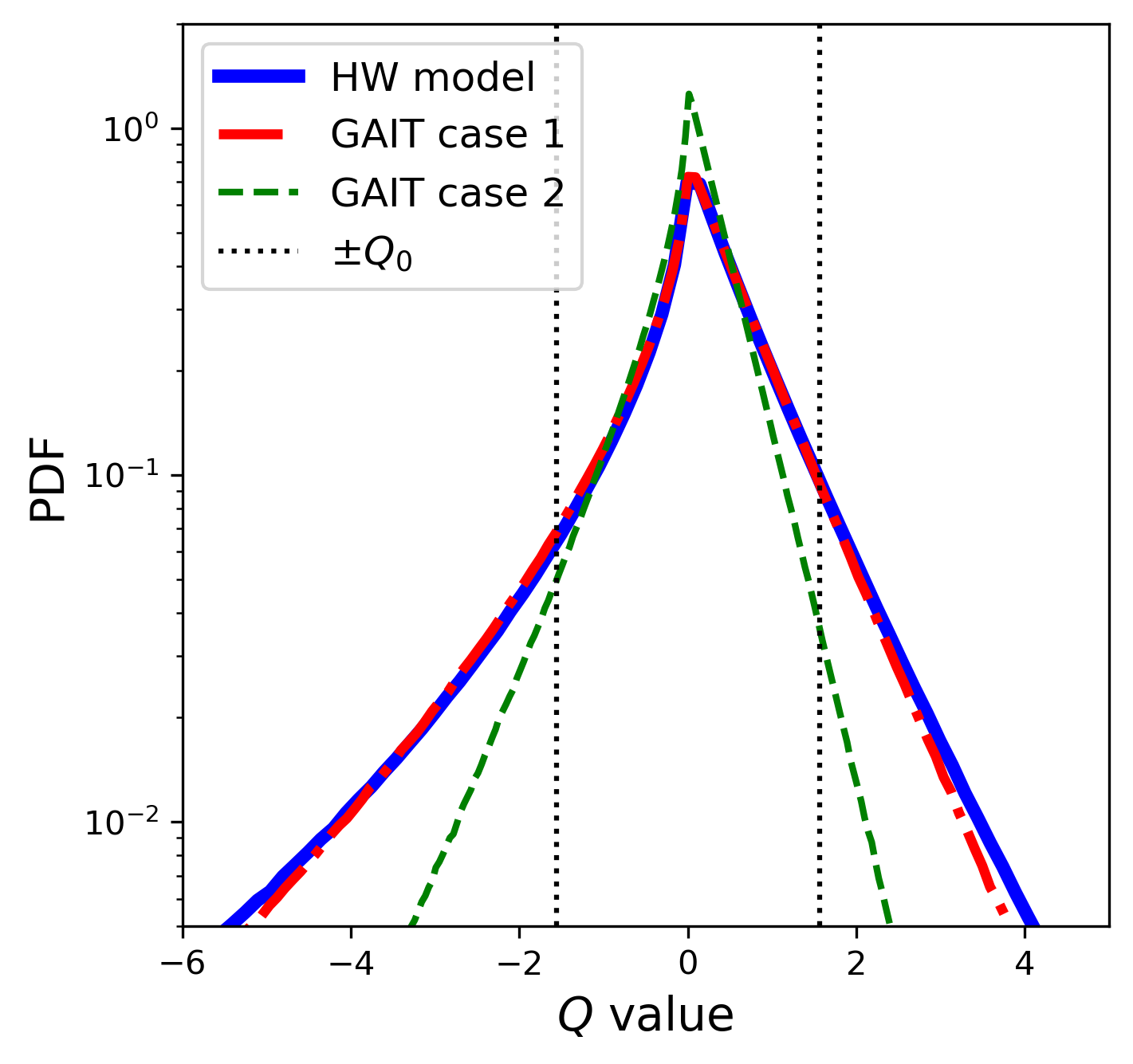}
    \caption{\raggedright \label{fig:ow_pdf} 
    Comparison between the HW and GAIT models flow topology.
    Top panel: spatial distribution of Okubo-Weiss field at a fixed time in HW (left), GAIT case 1 (center), and GAIT case 2 (right).
Bottom panel: probability distribution of $Q$ values in space and time. }
\end{figure}

A distinctive feature of 2D turbulence is the spontaneous formation and persistence of coherent structures, e.g. vortices and zonal flows \cite{sommeria1986experimental}. Understanding these structures is important because the trapping effect of vortices and the long displacements caused by zonal flows can have a critical role in transport, see for example \cite{del2000chaotic} and references therein. Accordingly, an important metric in the evaluation of the proposed AI model is how well the model captures the coherent structures of the flow. To quantify this, we use Proper Orthogonal Decomposition (POD) methods which have been extensively used in fluids \cite{berkooz1993proper} and plasmas \cite{beyer2000proper}. Contrary to Fourier spectral analysis, POD is based on data-driven empirical modes providing an optimal representation of the turbulence state in the energy norm. As discussed in \cite{Futatani2009}, the POD mode decomposition and spectra can be used to characterize the spatiotemporal coherence of HW plasma turbulence. 
Following this idea, the left panel of Fig.~\ref{fig:eulerian3} shows the POD singular values $w_l$ as a function of the rank, $l$, where $\phi(x,y,t)=\sum_{l=1}^N w_l u^{(l)}(x,y) v^{(l)}(t)$ is the singular value decomposition of the electrostatic potential. 
The close agreement, especially for \textit{case 1}, indicates that the GAIT model reproduces well the coherent structures of the HW model. Another metric of interest is the Eulerian correlation time of the electrostatic potential fluctuations, ${\cal C}(\tau)=\langle \overline {\phi(x,t) \phi(x,t+\tau) }\rangle$. The right panel in Fig.~\ref{fig:eulerian3} compares auto-correlation functions of the HW and the GAIT models, showing a very good reproduction of the turbulent dynamics.

To further compare the GAIT and the HW turbulence models, we consider the Okubo-Weiss (OW) parameter, used in \cite{kadoch2022} to characterize the flow topology in HW plasma turbulence. The OW parameter is defined as $Q=s^2-\Omega^2$, where $\Omega=\partial_x V_y -\partial_y V_x$ is the vorticity, $s^2=s_1^2 + s_2^2$ is the deformation with $s_1=\partial_x V_x -\partial_y V_y$ and $s_2=\partial_x V_y + \partial_y V_x$ \cite{Elhmaïdi_Provenzale_Babiano_1993}. Here  ${\bf V}= -\nabla\phi \times \mathbf{e}_z=(V_x,V_y)= (-\partial_y \phi, \partial_x \phi)$.  
Given the threshold, $Q_0 = \sqrt{\langle \overline {Q^2}\rangle} $, the turbulent flow can then be partitioned into three topologically distinctive regions: strongly elliptic (vorticity dominated) $Q \leq -Q_0$, strongly hyperbolic (deformation dominated) $Q \ge Q_0$, and intermediate $-Q_0 < Q < Q_0$. From a transport perspective, strongly elliptic regions tend to reduce transport due to particle trapping and strongly hyperbolic regions tend to enhance mixing due to particle dispersion. 
The agreement between the GAIT and the HW models in the flow topology  of the turbulent field is shown in Fig.\ref{fig:ow_pdf}. The top panels compare the Okubo-Weiss three levels ($Q<-Q_0$, $-Q_0 < Q < Q_0$, $Q>Q_0$) decomposition of the turbulence field at a given time. The bottom panel compares the
probability distribution of  $Q$ values in space and time $\{ Q_{i,j}(t_k)\}$, where $i=1,\ldots 64$, $j=1,\ldots 64$ and $k=1, \ldots N_s$. 

For all the Eulerian metrics shown in Figs.3-5, the agreement between the GAIT model and the HW turbulence model is remarkably good for Case 1. This is to be expected since this case optimizes the reconstruction of $\phi$ and $\nabla \phi$. Case 2, optimizes only $\nabla \phi$, and despite a high latent space regularity, yields less accurate results.

\begin{figure}[h]
    \begin{subfigure}{0.23\textwidth}
        \centering
        \includegraphics[width=0.9\linewidth, trim=0 0 0 0, clip]{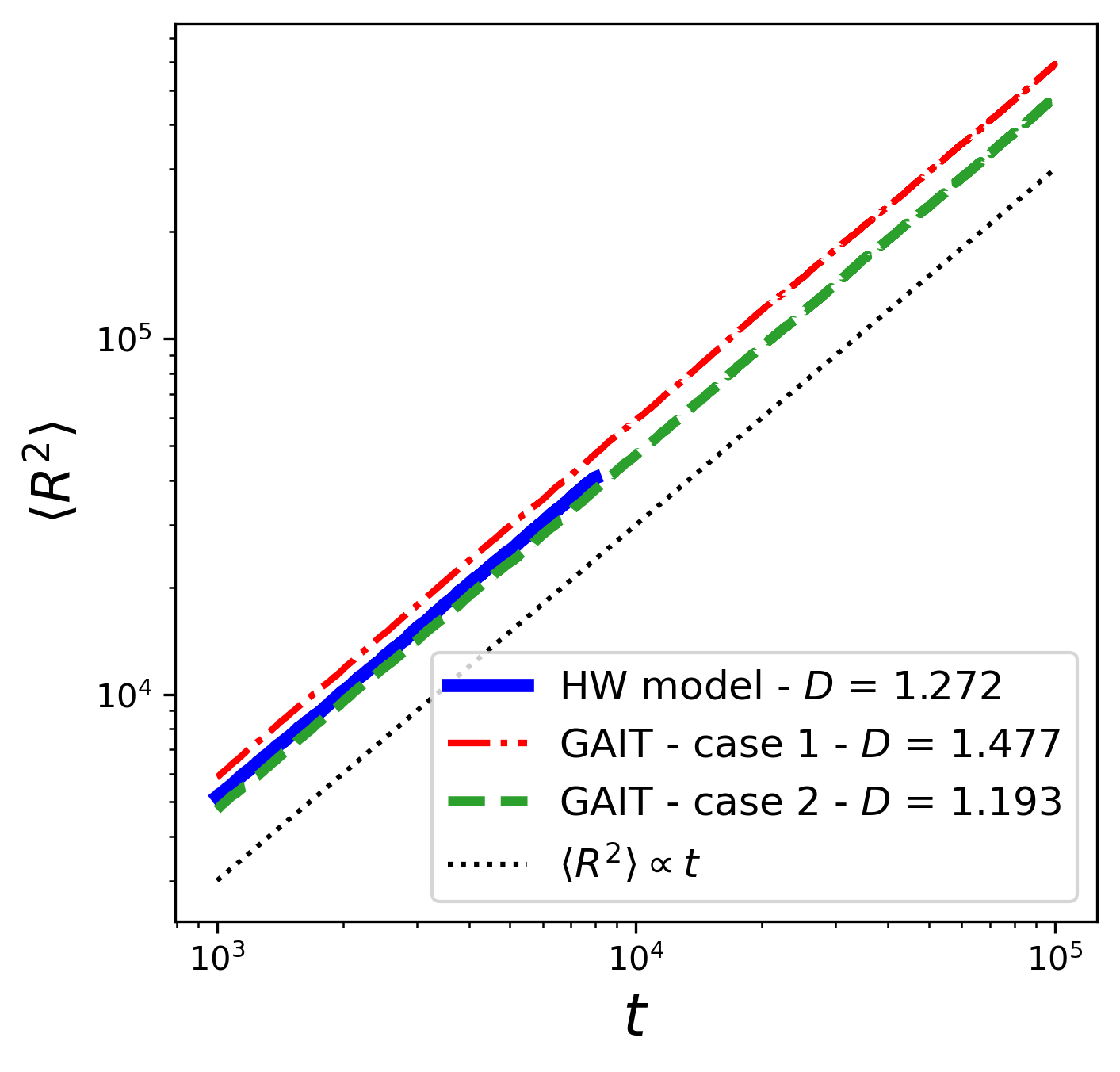}
        \label{fig:subfig5}
    \end{subfigure}
    \begin{subfigure}{0.24\textwidth}
        \centering
        \includegraphics[width=0.9\linewidth, trim=0 0 0 0, clip]{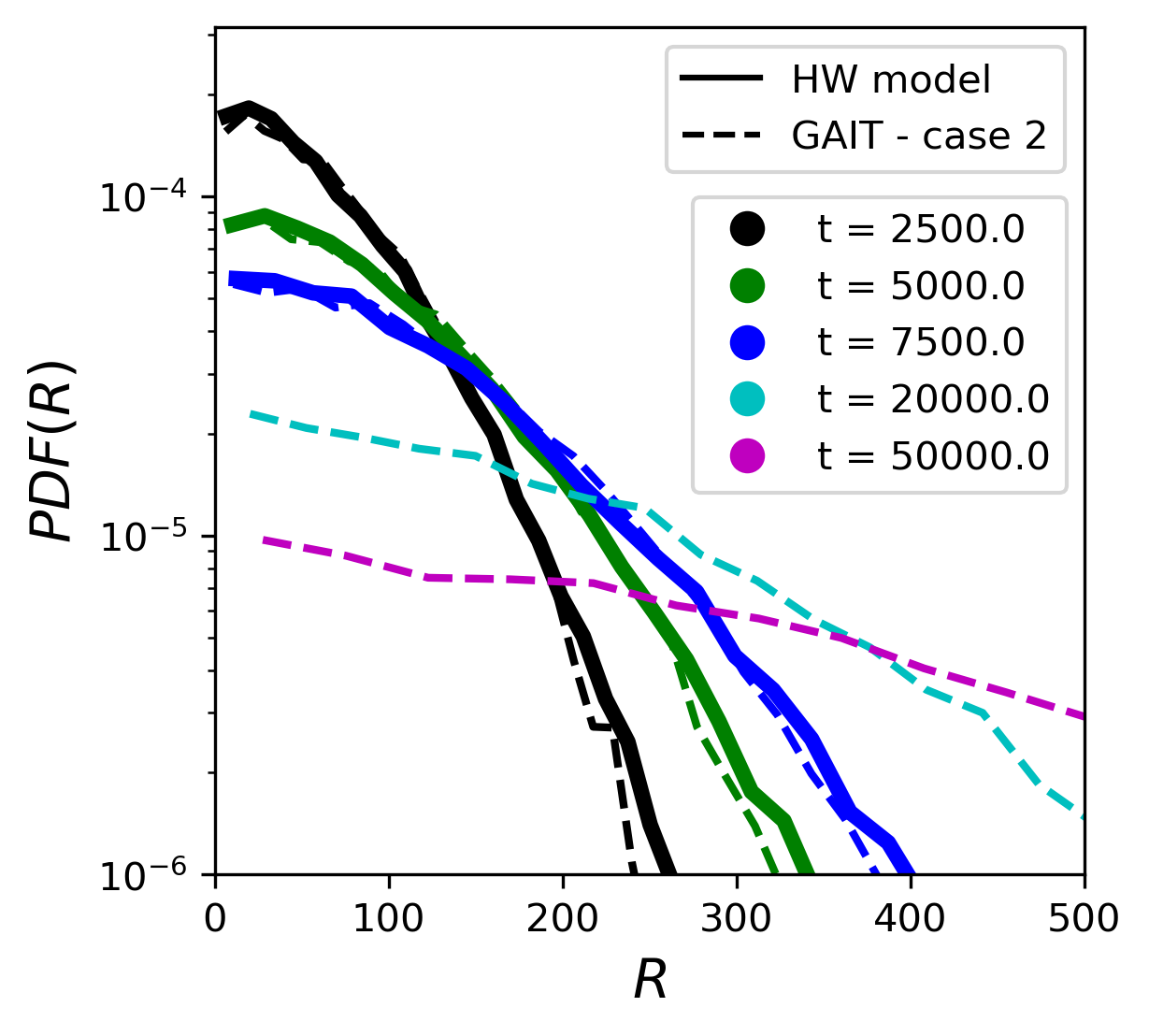}
        \label{fig:subfig6}
    \end{subfigure}
    \caption{\raggedright 
    Comparison of radial displacement statistics of tracers in the HW and GAIT models .
$\langle R^2 \rangle$ as function of $t$ in log-log scale (left). Probability distribution function of displacements at different times (right).}
    \label{fig:lagrangian1}
\end{figure}

As a final test, we study the turbulence-induced transport by solving the guiding center equations
$\frac{d\mathbf{x}}{dt} = -\nabla\phi \times \mathbf{e}_z$
for $10^4$ tracers initially distributed uniformly in the square domain
$\{ (x,y) | 0<x<2 \pi/k_0 \, , 0<y<2 \pi/k_0\}$.
The orbits are integrated using a Runge-Kutta method with a timestep $\Delta t = 0.2$. The data in the electrostatic potential snapshots was interpolated using linear interpolation in time and bilinear interpolation in space.
When computing the Lagrangian statistics the particle orbits are considered in the extended $\mathbb{R}^2$, i.e. the values of $(x_i(t),y_i(t))$ were not restricted to the 
$[0,2\pi/k_0] \times [0,2\pi/k_0]$ double periodic domain. 
The main quantities of interest are the PDF (Probability Distribution Function) of the displacement, $P(R)$, and the evolution of the mean squared displacement $\langle R^2(t)\rangle$ where
$ R = \| \delta \vec{r} - \langle\delta \vec{r}\rangle \|$, $\delta \vec{r}=\vec{r}(t)-\vec{r}(0)$ is the displacement vector and $\langle \cdot \rangle$ is an ensemble average. 
Figure \ref{fig:lagrangian1} shows that both the HW and the GAIT models exhibit  diffusive scaling $\langle R^2(t)\rangle \sim D t$, where $D$ is the diffusivity.

Case 2, that focuses on the optimization of $\nabla \phi$ (i.e., the ${\bf E}\times {\bf B}$ velocity) and high regularity of the latent space,  yields a better description of the Lagrangian transport with a diffusivity $D_{GAIT,2} = 1.19$ very close to the ground truth diffusivity $D_{HW} = 1.27$, along with a very good agreement on the PDFs of $R$. The smaller value of $w_{\mathbb{K}\mathbb{L}}$ in Case 1, reduces the latent space regularity and leads to a slightly higher value of the diffusivity, $D_{GAIT,1} = 1.47$. Most importantly, as proof of the fidelity of the turbulence generation process, the GAIT model preserves the transport properties well beyond ($t=10^5$) the numerical integration time of the HW model ($t=8 \times 10^3$).

The direct numerical integration of the HW model for 10,000 snapshots  took 
50 hours on a V100 GPU node of the French Jean-Zay supercomputer. 
On the same node, the training of the CVAE took 1 hour, and the training of the RNN five minutes. The generation of 10,000 snapshots of turbulent states with the GAIT model took 7.5 min, about  50 hrs/7.5 min $\sim$ 400 times faster than the direct numerical integration of the HW model.

The HW model requires the integration of the potential, $\phi$, and density, $n$, coupled equations. An advantage of the GAIT model is that it can provide a surrogate model of $\phi$ only, for applications not requiring $n$, like the one discussed here. If needed, a direct generalization of the GAIT model can be used to construct surrogates for $n$ and $\phi$.

This work has been carried out within the framework of the EUROfusion Consortium, funded by the European Union via the Euratom Research and Training Programme
(Grant Agreement No 101052200 — EUROfusion).
Views and opinions expressed are however those of the author(s) only and do not necessarily reflect those of the European Union or the European Commission.
Neither the European Union nor the European Commission can be held responsible for them.
This work has received financial support from the AIM4EP project (ANR-21-CE30-0018), funded by the French National Research Agency (ANR), and from the Oak Ridge National Laboratory, managed by UT-Battelle, LLC, for the US Department of Energy under Contract No. DE-AC05-00OR22725. 
All the simulations and training of neural networks reported here were performed on HPC resources of IDRIS under the allocations 2021-A0100512455, 2022-AD010512455R1 and 2023-A0140514165 made by GENCI.
\bibliography{manuscript}
\end{document}